\documentclass[journal]{IEEEtran}
\ifCLASSINFOpdf

\else

\fi

\usepackage{graphicx,amsmath,amssymb,cite}
\usepackage{subfigure}
\usepackage{color}
\usepackage{amsfonts}
\usepackage{amsmath}
\usepackage{graphicx}
\usepackage{amssymb}
\usepackage{stfloats}
\usepackage{mathrsfs}
\usepackage{algorithm}
\usepackage{algorithmic}

\begin{document}

%\title{An efficient deep learning network structure for secure spatial modulation  receiver }
\title{A Deep-learning-based Joint Inference for Secure Spatial Modulation Receiver}

\author{Feng Shu, Lin Liu, Yumeng Zhang, Guiyang Xia, \\ Xiaoyu Liu, Jun Li, Shi Jin, and Jiangzhou Wang.}
\maketitle

\begin{abstract}
  As a green and secure wireless transmission way, secure spatial modulation (SM) is becoming a hot research area. Its basic idea is to exploit both the index of activated transmit antenna and amplitude phase modulation (APM) signal to carry messages, improve security, and save energy. In this paper, we reviewed its crucial techniques: transmit antenna selection (TAS), artificial noise (AN) projection, power allocation (PA), and joint detection at desired receiver. To achieve the optimal performance of maximum likelihood (ML) detector, a deep-neural-network  (DNN) joint detector is proposed to jointly infer the index of transmit antenna and signal constellation point with a lower-complexity. Here, each layer of DNN is redesigned to optimize the joint inference performance of two distinct types of information: transmit antenna index and signal constellation point. Simulation results show that the proposed DNN method performs
    3dB better than the conventional DNN structure and is close to ML detection in the low and medium signal-to-noise ratio regions in terms of the bit error rate (BER) performance, but its complexity is far lower-complexity compared to ML. Finally, three key techniques TAS, PA, and AN projection at transmitter can be combined to make SM a true secure modulation.
\end{abstract}

%\begin{IEEEkeywords}
%  Machine learning, deep neural network(DNN), spatial modulation, secrecy rate, artificial noise, power allocation
%\end{IEEEkeywords}

\section{Secure spatial modulation and deep learning}
  Spatial modulation (SM) concept was first proposed by  Chau and Yu in \cite{Space_modulation2001}. They had creatively proposed the concept of SM: carry additive bit information via antenna indices. In \cite{Mesleh2008Spatial}, the authors made a systematic and in-depth investigation of SM, and officially named it as SM. At the same time, the basic principle of SM was also explained. SM exploits both the index of activated transmit antenna and amplitude phase modulation (APM) signal to carry messages. Compared to Bell Laboratories Layer Space-Time (BLAST) and space time coding (STC), SM system achieves a good balance between spatial multiplexing and diversity. We call it as the third way between BLAST and STC. Compared to BLAST and STC,   SM has a good advantage of high energy efficiency (EE) due to the use of less active RF chains. Thus, it is a green wireless transmission technique.

  Wireless communication is usually prone to passive eavesdropping and active malicious attacks due to its broadcast characteristics. Although there is a series of mature encryption algorithms in the upper layer of network protocol, it is still fragile in wireless communication. To address this issue, the physical layer security (PLS) technology becomes a nature choice, and   enhances its security from the perspective of information theory. PLS has been extensively studied in \cite{PLSchallenges2015}. PLS will work with traditional cryptography to play a key role and provide an incremental  guarantee for the future personal privacy protection and information network security. Working with encryption together, a dual protection of transmitting confidential messages (CMs) will be achieved.

  In the past decade, secure modulation emerges as an special form of multiple-input-multiple-output (MIMO). It mainly consists of two types: directional modulation (DM) and secure SM(SSM). DM using beamforming with the help of artificial noise (AN) can securely deliver confidential messages (CMs) to desired users in line-of-sight channel, and  is unsuitable for fading channels. Instead, SM is naturally suitable for fading channel.

  By introducing security into the SM, it is able to transmit CMs over the fading channel. However, transmitting CMs via SM is an attractive and very important issue \cite{Feng2018Two} \cite{shu2018secure} \cite{xia2019antenna}.  In \cite{shu2018secure}, the authors  made an extensive investigation of TAS methods in SSM systems. Then, two high-performance  transmission antenna selection schemes: leakage-based and  maximum SR, have been proposed to improve the SR performance, and the generalized Euclidean distance-optimized antenna selection method has been generalized to provide a secure transmission.  In \cite{xia2019antenna}, an active antenna-group (AAG) selection is proposed to maximize the average secrecy rate (SR) in the case of limited active antenna pattern and finite-alphabet inputs.

   In SSM,  how to optimize and design the AN projection matrix  has a substantial impact on the SR performance. In \cite{Wu2015Secret, xia2019antenna}, AN was projected  onto the null-space of the desired channel to improve the security of communication systems. The main benefit of this scheme is that the AN projection matrix has a closed-form expression. However, such a scheme might cause some secrecy performance loss due to lack a holistic consideration of secure communication systems. In other words, the design of the beamformer of AN in conventional way is merely achieved from the perspective of the desired receiver.

\begin{figure*}[htbp]
%\begin{figure*}[h]
\centering
\includegraphics[width=1.0\textwidth]{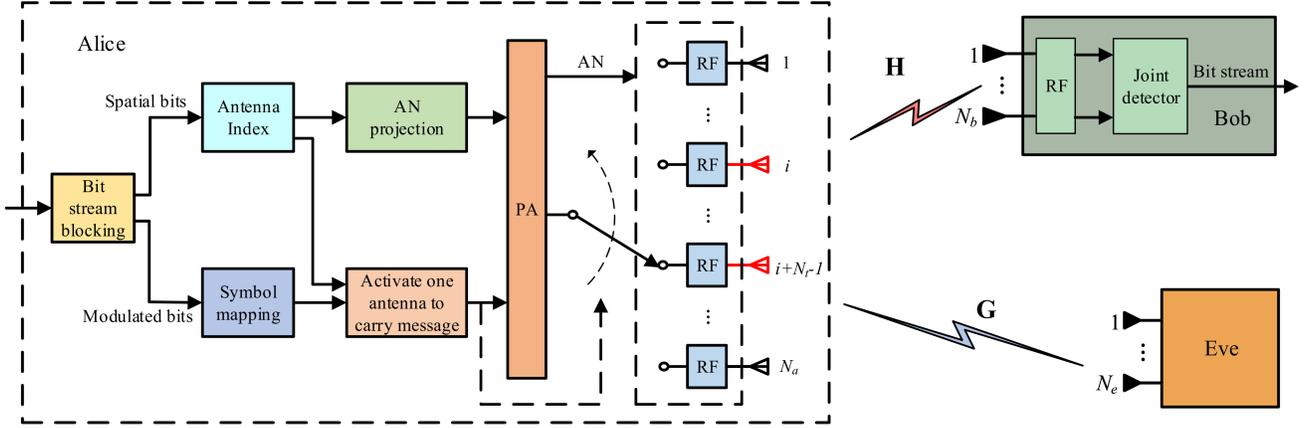}
\centering
\caption{Schematic diagram of secure SM network.}\label{Spatial modulation_fig1}
\end{figure*}

  Intelligent communication is considered as one of the mainstream directions of the follow-up development of mobile communication after 5G. The basic idea is to introduce intelligent elements into all layers of mobile communication system, realize the organic integration of mobile communication and artificial intelligence technology, and greatly improve the efficiency of mobile communication system.  The previous research results have focused on the application layer and the network layer, and the main idea is to introduce machine learning, especially deep learning (DL), into wireless resource management, channel decoding, and other fields.
 % Machine learning is a multidisciplinary subject involving many disciplines such as probability theory, statistics, approximation theory, convex analysis, and algorithm complexity theory. Common machine learning algorithms include regression algorithm, instance-based algorithm, decision tree algorithm, Bayesian algorithm, clustering algorithm, deep learning and neural network algorithm. Machine learning has been widely used in the upper layers of wireless communication systems, such as cognitive radio, wireless network resource allocation, and the like. However, the traditional machine learning algorithm is not well applied due to the limitation of complex channel environment and learning ability in the application of the physical layer.
  DL is one of the most important breakthroughs in the field of artificial intelligence in the past decade. The author details the DL algorithms in \cite{Schmidhuber2015Deep}. It has been successfully applied in many fields such as computer vision, natural language processing, speech recognition, etc., and has achieved great success. Due to the new features of future communication, such as complex scenes with unknown channel models, high-speed and accurate processing requirements, many scholars have introduced DL into the physical layer of wireless communication \cite{C.Jiang2017Machine}. In the physical layer, there is a new trend of combining wireless transmission and DL.

  In \cite{He2018Deep}, the authors considered channel estimation for millimeter-wave massive MIMO systems. An approximate messaging network based on learning denoising was proposed for channel estimation, which can learn channel structure and estimate channel from a large amount of training data. In \cite{Nachmani2016Learning}  a new framework were proposed for integrating large-scale MIMO and DL to address the problem of channel estimation and DOA estimation \cite{Hongji}. Deep neural network (DNN) is used for offline learning and online learning, and the statistical characteristics of the wireless channel and the spatial characteristics of the angle domain are effectively learned. A novel deep learning assisted sparse coded multiple access scheme is proposed in\cite{ Kim2018Deep}. By using a DNN-based encoder and decoder adaptively construct a codebook that minimizes the bit error rate in \cite{nachmani2018deep}.

  In Fig.~\ref{Spatial modulation_fig1}, a typical secure SM system  is shown. In this figure, four main tools including transmit antenna selection (TAS), beamforming of confidential messages, AN projection,  and power allocation (PA) are fully utilized to achieve a SSM. In such a network, at desired receiver, the joint detection of transmit antenna index and signal constellation point is required. Joint detection performance is very important. Conventionally, the optimal maximum likelihood (ML) detector is a natural choice. But, as the number of transmit antennas tends to medium-scale or large-scale or the size of signal constellation goes to medium-scale  or large-scale, the ML joint detector is confronted with a complexity bottleneck, i.e., exponential complexity. To reduce the computational complexity of receiver, the DNN-based joint detector is proposed to jointly infer the transmit index and signal constellation point in this paper. Compared to ML, the joint DL detector is low-complexity and reach the optimal performance of ML.

  \section{System model and Transmit antenna selection}

  Consider a typical SSM system as shown in Fig.~\ref{Spatial modulation_fig1}. In this system, there is a transmitter (Alice) equipped with $ N_a $ transmit antennas. Without loss of generality, when the number of antennas at the transmitter is not a power of two,  $ N_t = 2^{\left\lfloor\log _{2}^{N_a}\right\rfloor}$, out of $ N_a $  transmit antennas are selected for mapping the bits to the antenna index. The $\log_{2}^{M} $ bits are used to form a constellation symbol, where $M$ is the signal constellation size. As a result, the spectral efficiency is $\log _{2} N_{t}+\log _{2} M $ bits per channel use (bpcu).

  Referring to the secure SM model in \cite{shu2019high}, the transmit signal vector with the aid of AN can be given by
\begin{equation}
\mathbf{x}=\sqrt{\beta P_{S}} \mathbf{e}_{n} \mathbf{s}_{m}+ \sqrt{(1-\beta)P_{S}} \mathbf{P}_{AN}\mathbf{n}
\label{transmit signal }
\end{equation}
 where $P_{S}$ denotes the total transmit power constraint and $\beta$ is the PA factor. $\mathbf{e}_{n} $ is the $n$-th column of identity matrix $\mathbf{I}_{Nt}$, and $s_{m}$ is the digital constellation symbol with a normalized power $\emph{E}[|s_m|^2] = 1$. Additionally, $\mathbf{P}_{AN}$ is the AN projection matrix and $\mathbf{n}\in \mathbb{C}^{N_{t} \times 1} $ is the corresponding AN vector. Then, the receive signal at desired receiver Bob  can be formulated as follows
 \begin{equation}
 \mathbf{y}_{d}= \sqrt{\beta P_{S}}\mathbf{HT}_{k} \mathbf{e}_{n} s_{m}+\sqrt{(1-\beta )P_{S}}\mathbf{H} \mathbf{T}_{k} \mathbf{P}_{A N} \mathbf{n}+\mathbf{n}_{b}
 \end{equation}
 %\begin{equation}
% \mathbf{y}_{e}=\sqrt{\beta P_{S}}\mathbf{G} \mathbf{T}_{k} \mathbf{e}_{n} s_{m}+\sqrt{(1-\beta)P_{S}}\mathbf{G} \mathbf{T}_{k} \mathbf{P}_{A N} \mathbf{n}+\mathbf{n}_{e}
%\end{equation}
where $\mathbf{H}$ are the complex channel gain matrix from Alice to Bob and to Eve, $\mathbf{T}_{k}$ is transmit antenna selection matrix.

%In contrast, the learning-based detection of $\hat{n} \hat{m}$ can be formulated as
%\begin{equation}
%[\hat{n}, \hat{m}]=\underset{n \in\left[1, N_{t}\right], m \in[1, M]}{\arg \min }\left\|\mathbf{y}_{b}-\sqrt{\beta P_{S}} \mathbf{H} \mathbf{T}_{k} \mathbf{e}_{n} s_{m}\right\|^{2}
%\end{equation}
%\subsection{Transmit antenna selection}
  Selecting an active antenna group can be adopted to further improve the performance of SM systems. There are several existing TAS methods for secure SM system as follows: random, leakage \cite{Feng2018Two}, and  generalized Euclidean distance antenna selection (EDAS). For the leakage-based TAS strategies,  the signal-to-leakage-and-noise ratio (SLNR) of CM from each transmit antenna is computed and formed a sequence  of SLNRs, where SLNR is defined as the ratio of the receive signal power at Bob to the sum of the receive power of CM at Eve, receive AN power, and channel noise variance.  Then a low-complexity sorting algorithm places  the values of SLNR in decreasing order. The antennas corresponding  antennas associated with the  top $N$ SLNRs is chosen, called Max-SLNR \cite{Feng2018Two}. The Max-SLNR can achieve the near-optimal SR performance with a low-complexity.

  From the aspect of decoding performance at receiver, generalized EDAS performs best in terms of bit error rate. The generalized EDAS Method is aim to select a TAS pattern of maximizing the minimum Euclidean distance over desired channel or minimizing the minimum Euclidean distance over eavesdropping channel  due to the fact that the minimum distance has a direct relationship to BER.

%\end{enumerate}

\begin{figure}[htbp]
\centering
\includegraphics[width=0.5\textwidth]{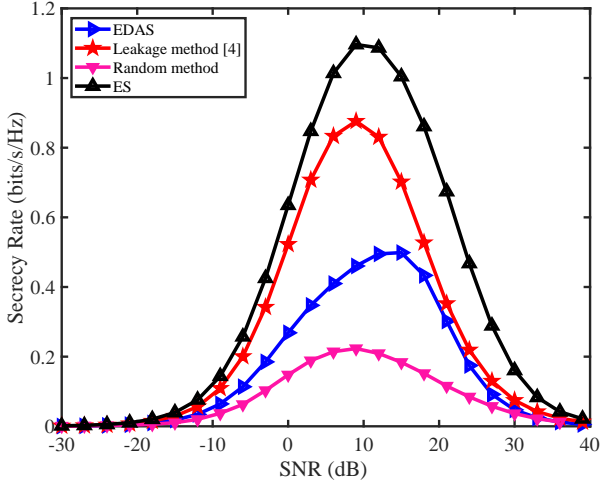}
\caption{Comparison of SR performance of various TAS methods.}\label{Comparison-of-SR-No-AN}
\end{figure}
 Fig.~\ref{Comparison-of-SR-No-AN} demonstrates the SR performance comparison of the optimal exhaustive search  (ES), Max-SLNR, EDAS, and random methods without the aid of AN. From this figure, it is seen that  the four methods have a decreasing order in SR performance as follows: ES,  Max-SLNR, EDAS, and random method. Additionally, we also find an interesting result: all the SR curves first go up as hills, then reach their peaks, and finally go down hills as SNR increases. In other words, all the SR curves have main peaks, and can be approximately viewed as concave functions of the SNR.

\begin{figure}[htbp]
\centering
\includegraphics[width=0.5\textwidth]{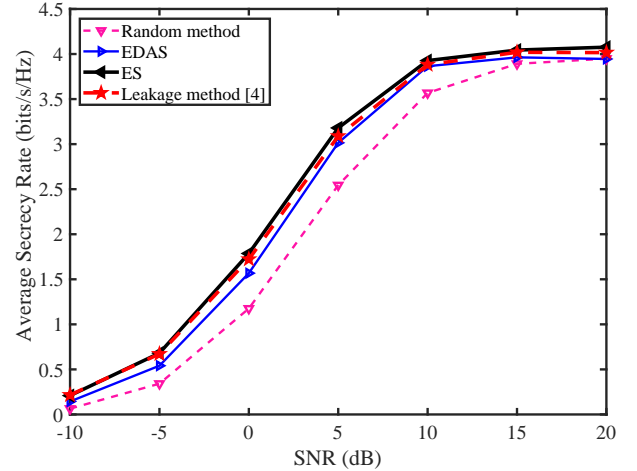}

\caption{Comparison of SR performance of various PA strategies.}\label{Comparison-of-SR-With-AN}
\end{figure}

 Fig.~\ref{Comparison-of-SR-With-AN} makes a SR performance comparison of ES,  Max-SLNR, EDAS, and random methods with the aid of AN. From this figure, it is seen that  the four methods have still an decreasing order in SR as follows: ES,  Max-SLNR, EDAS, and random. In particular, observing Fig.~\ref{Comparison-of-SR-No-AN} and Fig.~\ref{Comparison-of-SR-With-AN},  we find an important fact: with the aid of AN, the SR performance can be improved significantly, especially in the medium and high SNR regions. The values of SR for the four methods grows gradually as SNR increases. When SNR enters the high SNR region, their SR performance  reaches their corresponding SR ceils.

 %T%he SR of the three methods is close to zero at low SNR. the reason is that desired user can't decode correctly in the low SNR region. However, when SNR is in high region, the SR is close to zero, The main reason is that as the SNR increased the eavesdropper can efficiently obtain confidential messages. The transmission rates of desired user and eavesdropper are close to spectral efficiency $\log _{2} N_{t}+\log _{2} M $, Thus the SR is close zero.  To improve the security performance, artificial noise can be introduced to interfere the eavesdropper. However, the power allocation between useful signal and artificial noise must be considered when artificial noise is introduced.

%% To address this problem, we can Allocate part of the power to artificial noise. Next, discussed artificial noise design and power distribution.
\section{Beamforming, AN projection, and Power allocation}
 Since secure SM channel can be viewed as a discrete-input continuous-output memoryless channels (DCMC), it is very hard to find a closed-form expression for mutual information in such a network. Mutual information contains the expected items of noise, usually with high computational complexity. Only in the case that the input is a Gaussian signal, The SR expression will have a very concise form.

 For DCMC, in general,  an approximate estimation SR expression is used instead of exact SR, it is difficult to convert to a convex problem. There is an inversion in the expression, and the outer layer needs to solve the expectation. This is a rather complicated issue.  In addition, it is difficult for the transmitter to optimize the CM beamforming vector and AN projection matrix by maximizing SR.   This is a  a challenging problem in the coming future. However, at the cost of some SR performance loss, some low-complexity and closed-form methods can be used. For example, the AN projection matrix can be  constructed from the null-space of the desired channel from Alice to Bob while the CM beamforming vector is also formed from the null-space of the desired channel from Alice to Eve. If you want to further the SR performance, the leakage-based rule is used to optimize the design of AN projection matrix and CM beamforming vector.
%Considered finite-alphabet signal inputs, the mutual information of Bob can be formulated as follow
%\begin{equation}
%\begin{aligned}& I\left(\mathbf{x};\mathbf{y}_{b}| \mathbf{H}, \mathbf{T}_{k}\right) = \log _{2}{M N_{t}}- \\ &\frac{1}{M N_{t}} \sum_{i=1}^{N_{t}M} \mathbb{E}_{\mathbf{n}_{b}^{\prime}} \left\{\log _{2} \sum_{j=1}^{N_{t}M} \exp\left(-f_{b, i, j}+\left\|\mathbf{n}_{b}^{\prime}\right\|^{2}\right) \right\}\end{aligned}
%\end{equation}
%where $f_{b, i, j}=\left\|\sqrt{\beta Ps}\mathbf{W}_{b}^{-1/2} \mathbf{H}\mathbf{T}_{k} \mathbf{d}_{i j}+\mathbf{n}_{b}^{\prime}\right\|^{2}$. Here, $\mathbf{d}_{i j}=\mathbf{x}_{i}-\mathbf{x}_{j}$,$\mathbf{x}_{i}-\mathbf{x}_{j}$ is one of possible transmit vectors in the set of combining antenna and all possible symbols. $\mathbf{W}_{b}$ is the covariance matrix of interference plus noise of Bob, where $\mathbf{W}_{b}=(1-\beta)Ps\mathbf{H}\mathbf{T}_{k} \mathbf{P}_{AN}\mathbf{P}_{A N}^{H} \mathbf{T}_{k}^{H} \mathbf{H}^{H}+\sigma_{b}^{2} \mathbf{I}_{N_{b}} $ .
%$\mathbf{W}_{b}^{-1/2}$ is a whitening filter matrix that can change color noise plus AN to white noise. $\mathbf{n}_{b}^{\prime}=\mathbf{W}_{b}^{-1/2}\left(\sqrt{(1-\beta) Ps} \mathbf{H}\mathbf{T}_{k}\mathbf{P}_{A N} \mathbf{n}+\mathbf{n}_{b}\right)$

%In order to improve the security performance, the authors in \cite{Wu2015Secret} studies the security precoding spatial modulation system, and the authors in \cite{Liu2017Secure} proposes a full-duplex receiver, which sends interference signals to prevent the eavesdropper from eavesdropping while receiving signals.

PA, as an efficient way to enhance security, has been investigated in  \cite {shu2019high}. By adjusting the PA factor, the power can be allocated between CM and AN freely to affect the SR and BER performance. By simulation and proof, it confirms that SR is a concave function of the PA factor $\beta$. Although exhaustive search (ES) can be employed to search  the optimal PA factor, its computational complexity is high. Thus, in \cite {shu2019high}, a novel PA strategy, called Max-P-SINR-ANSNR where 'P' is short for product, and 'ANSNR' stands for AN-to-signal-plus-noise ratio, presented a closed-form expression for the PA factor. This dramatically reduces the complexity of ES.
%%put the power allocation is a fixed value
\begin{figure}[htbp]
\centering
\includegraphics[width=0.5\textwidth]{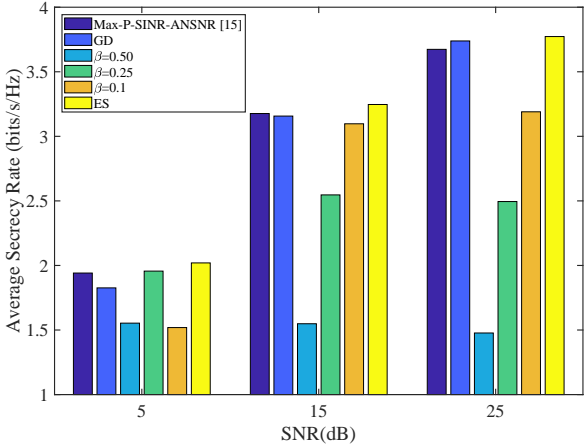}
\caption{Comparison of SR of various PA strategies.}\label{Comparison-of-PA}
\end{figure}

 Fig.~\ref{Comparison-of-PA} makes a comparison of several typical PA strategies: ES, fixed, gradient descent (GD),  and Max-P-SINR-ANSNR.  From this figure, it follows that the Max-P-SINR-ANSNR and GB strategies in \cite{shu2019high} are close to the optimal ES, but their complexity is dramatically lower than ES.  Comparing the three methods with fixed power allocation factors, it can be seen that the SR at $\beta=0.5$ is the lowest one, and $\beta=0.1$ is the highest one in the value of high SNR. This is because when the SNR is high, both Bob and Eve have a very good quality of channel, and a large portion of transmit power may be allocated to AN to disturb eavesdropper, so as to obtain a high security performance. Additionally, due to its closed-form expression of Max-P-SINR-ANSNR, it strikes a good balance between complexity and performance.

\section{Proposed DNN-based Joint Inference of antenna index and constellation point}
 Assuming Bob has the perfect channel state information (CSI) of $\mathbf{H}$ and $\mathbf{T}_{k}$, the joint ML detector (MLD) at desired receiver can be casted as follows
\begin{equation}
[\hat{n}, \hat{m}]=\underset{n \in\left[1, N_{t}\right], m \in[1, M]}{\arg \min }\left\|\mathbf{y}_{d}-\sqrt{\beta P_{S}} \mathbf{H} \mathbf{T}_{k} \mathbf{e}_{n} s_{m}\right\|^{2}_{2}
\end{equation}
where $\hat{n}$ and $\hat{m}$ denote the index of transmit antenna and signal constellation point by joint MLD, respectively. The MLD requires the computational complexity $N_{TAS}\times N_{SC}$ floating-point operations (FLOPs) where $N_{TAS}$ and $N_{SC}$ denote the complexity of TAS scheme and the constellation size, respectively. Obviously, this complexity is a product. As $N_{TAS}$ , $N_{SC}$ or both tends to large-scale, the total complexity will become a large number.

\begin{figure*}[htbp]
\centering
\includegraphics[width=1.0\textwidth, angle=0]{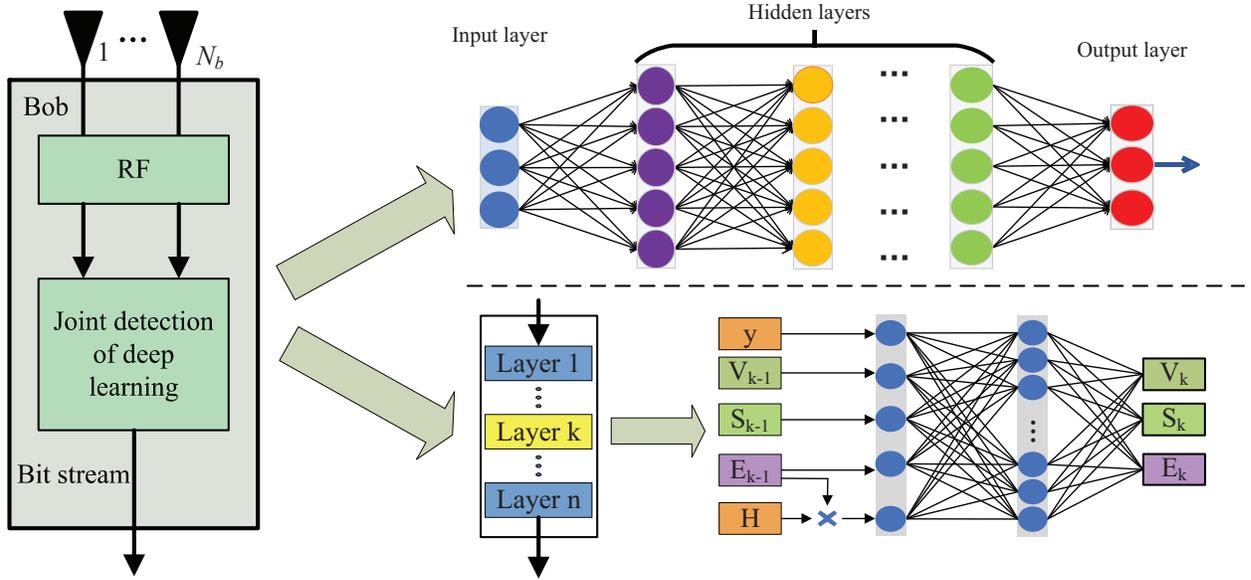}
\caption{Proposed novel DNN structure for joint inference of antenna index and constellation point.}\label{dnn_fig01}
\end{figure*}

To reduce this complexity, a DNN is a natural choice due to the fact that determining which antenna or constellation point is actually  a kind of classifying. However, the conventional DNN structure shown in Fig.~\ref{dnn_fig01}, locating on the right-upper corner, is verified to be 3dB worse than the ML in terms of BER given a fix BER=$10^{-3}$. To completely remove the 3dB performance gap, a novel DNN structure shown in the left-bottom corner of Fig.~\ref{dnn_fig01} is proposed. Here,  each layer is redesigned to have three kinds of outputs: $\mathbf{S}_{k}, \mathbf{E}_k$ and $\mathbf{V}_k$, where $\mathbf{S}_{k}$ and $\mathbf{E}_k$ are the estimate of constellation symbol and antenna index in the $k$-layer, respectively. $\mathbf{V}_k$ is the hidden output vector of the $k$-layer, and is also the input of the next layer. The DNN-based joint inference idea is to minimize the loss function
 \begin{align}
\|\mathbf{x}-\hat{\mathbf{x}}(\mathbf{y}_{d},\mathbf{H})\|^2_{2}=\|\mathbf{x}-\emph{F}(\mathbf{y}_{d},\mathbf{H};\mathbf{w},\mathbf{b})\|^2_{2}
\label{loss function}
\end{align}
by random GD and back-propagation methods, where
\begin{align}
\emph{F}(\mathbf{y}_{d},\mathbf{H};\mathbf{W},\mathbf{b})=\mathbf{f}^{n-1} \bigg(\mathbf{W}^{n-1}\mathbf{f}^{n-2} \bigg(\cdots \bigg(\mathbf{W}^1\mathbf{f}^{1}\bigg(\nonumber\\ \mathbf{W}^0
\left(\begin{array}{c}\mathbf{y}_{d} \\ \mathbf{Vec(H)}\end{array} \right)+\mathbf{b}^0 \bigg)+\cdots \bigg)+\mathbf{b}^{n-1} \bigg)
\end{align},
 where $\mathbf{W}^k$ is the  matrix of weight coefficients corresponding to layer $k$,  and $\mathbf{b}^k$ stands for the bias vector for the corresponding layer. $\mathbf{f}^{(k)} (\cdot) $ represents nonlinear activation function of layer $k$ and describes an input-output mapping. $\mathbf{Vec}(\mathbf{H})$ is the vectorization of the matrix $\mathbf{H}$

To solve the gradient disappearance problem, some new activation functions are adopted to replace the classical sigmoid activation function. Among them, the Rectified Linear Unit (ReLU) is commonly used. Introducing nonlinearity by zeroing a function value less than zero. When the input value is greater than zero, the function is a linear function, which simplifies the gradient calculation and the gradient value will not decrease with the increase of the input. This alleviates the gradient disappearance problem.

To speed up the convergence and reduce the computational complexity, the classical DG algorithm adjusts to the fall of the stochastic gradient. that is, randomly selects a sample to calculate the loss function and the gradient. However, the random selection of sample selection causes a large volatility in the training process, that is, the network cannot converge to the optimal solution, but fluctuates around the optimal solution. Take a compromise, using a batch gradient descent method, each time a batch of samples is selected to calculate the loss function and gradient. It's also not guaranteed to be globally optimal by this method,  but we can receive this solution as the loss function is reduced to be small.

Even using the above two strategies, neural networks are not very easy to train. Then, some of the current popular neural network training techniques can be adopted to improve its performance, speed and stability. The use of a moving average model enhances the stability of the parameters. Dropping out units (both hidden and visible) by a probability when training neural networks can efficiently address the problem of network over-fitting and improve its generalization ability. Additionally, the batch normalization is applied to improve the performance.

Now, we present numerical results to validate the effectiveness of our proposed DNN-based scheme. Constructing training and validation sets: generating a large amount of training data by performing baseband numerical simulation on the SM-MIMO channel model, and performing two-dimensional labeling of the antenna and the constellation point on the received vector data.  The structure of the neural network depth and the number of neurons per layer are determined by initializing the weight matrix and bias of the network. Then hyperparameters: maximum number of training, batch size, initial learning rate, and drop probability, are also initialized. The network is trained by training set, and the verification set is used to verify the network after every 100 iterations to decide whether to terminate early. Apply the above learning completion network to a desired receiver of SM to verify its BER  performance.
\begin{figure}[htbp]
\centering
\includegraphics[width=0.5\textwidth]{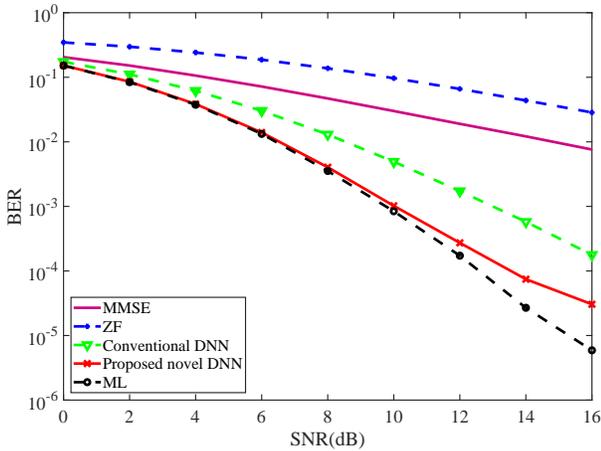}
\caption{BER performance comparison among proposed DNN, joint ML,  conventional DNN, ZF, and MMSE with $N_a$=4, $N_r$=4 and QPSK.}\label{fig2QAM44}
\end{figure}

 Fig.~\ref{fig2QAM44} illustrates the curves of average BER versus SNR for the five methods with $N_t=4, N_r=4$, and modulation being QPSK, where  several classic detector such as ML,  zero-forcing, minimum mean square error (MMSE) are used as performance benchmarks.  From Fig.~\ref{fig2QAM44}, it is seen that  the proposed novel DNN-based method approaches the ML method when SNR is lower than 10dB, and much better than ZF, MMSE and conventional DNN ones in almost all SNR regions in terms of BER performance. It is evident that the conventional DNN-based method is 3dB worse than the proposed novel DNN  and ML ones at BER$=10^{-3}$. In summary, the network's performance can be significantly improved by adjusting the structure of DNN.
 % adjusting the structure of DNN can significantly improve the network performance.
\section{Open problems}
 There are still many open problems existed to be addressed. Here, we list several important ones of them as follows:
 \begin{enumerate}
  \item  As the number of antennas at SM transmitter tends to large-scale, the circuit cost and complexity becomes a significant obstacle for practical applications of SM. A hybrid analog-digital  structure is preferred. In such a scenario,  how to achieve an optimal strategy of transmit antenna subarray is challegening problem. This problem can be modelled an integer optimization problem. The key is to develop a low-complexity algorithm.
      % A hybrid analog and digital (HAD) structure structure is an inevitable alternative. Sub-array structure % in HAD can result in a dramatic reduction in circuit cost and complexity. However, % in such a structure, there are open problems of how to offer high-precision SR measurement and robust TAS scheme for SM.

  \item If Eve works in a full-duplex model and becomes  an active eavesdropper, i.e., jamming, how to optimize the design of the transmitter at Alice in order to reduce the effect of jamming from Eve and at the same time achieve a feasible performance is a hard task. For Bob, the receive beamforming scheme of combating the jamming is preferred.

  \item  By adjusting the neural network structure of deep learning, the BER performance of the proposed structure  with QPSK is close to the performance of ML detection. To extend it to the high-order modulation is still an open problem. This requires us to optimize the structure of deep learning.

  \item In the presence of CSI measurement errors, how to construct robust beamforming, PA, and TAS by taking the statistical property of CSI error into account requires a great effort. In particular, the first task is to formulate the SR expression or approximate expression in such a scenario. This will pave a way for robust beamforming, PA, and TAS.
  \item  Due to non-convexity of the SR expression, AN optimization is a challenging task because the beamformer of AN is always included in a complicated expression, especially when a rough/statistic CSI of Eve's channel is available at the transmitter.

  \item  As the number of TAs further increases, how to optimize an AAG for enhancing the security of the communication systems becomes intractable, this is because the computational complexity of evaluating the accurate SR will be explosively grown upon increasing the number of TAs. That requires us exploring some concise metrics instead of the original function to evaluate the value of SR.

\end{enumerate}

\section{Conclusion}
In this article,  the great potential of secure SM has been highlighted as a key secure tool for future vehicular communications, IoT, UAV, smart transportation, and satellite communications.  We review its key techniques: TAS schemes, PA strategies, and joint detection methods at desired receiver. A new DNN structure was proposed to jointly infer the transmit antenna index and signal constellation point, i.e., a new joint detection way with low-complexity.  Also, we have raised several new open important future research problems. Finally, in our view, secure SM will have  wide diverse promising applications in the near future.

\ifCLASSOPTIONcaptionsoff
  \newpage
\fi
\bibliography{IEEEfull,reference}

\begin{thebibliography}{10}
\providecommand{\url}[1]{#1}
\csname url@samestyle\endcsname
\providecommand{\newblock}{\relax}
\providecommand{\bibinfo}[2]{#2}
\providecommand{\BIBentrySTDinterwordspacing}{\spaceskip=0pt\relax}
\providecommand{\BIBentryALTinterwordstretchfactor}{4}
\providecommand{\BIBentryALTinterwordspacing}{\spaceskip=\fontdimen2\font plus
\BIBentryALTinterwordstretchfactor\fontdimen3\font minus
  \fontdimen4\font\relax}
\providecommand{\BIBforeignlanguage}[2]{{%
\expandafter\ifx\csname l@#1\endcsname\relax
\typeout{** WARNING: IEEEtran.bst: No hyphenation pattern has been}%
\typeout{** loaded for the language `#1'. Using the pattern for}%
\typeout{** the default language instead.}%
\else
\language=\csname l@#1\endcsname
\fi
#2}}
\providecommand{\BIBdecl}{\relax}
\BIBdecl

\bibitem{Space_modulation2001}
Y.~A. {Chau} and {Shi-Hong Yu}, ``Space modulation on wireless fading
  channels,'' in \emph{IEEE 54th Vehicular Technology Conference. VTC Fall
  2001. Proceedings (Cat. No.01CH37211)}, vol.~3, Oct 2001, pp. 1668--1671
  vol.3.

\bibitem{Mesleh2008Spatial}
R.~Y. {Mesleh}, H.~{Haas}, S.~{Sinanovic}, C.~W. {Ahn}, and S.~{Yun}, ``Spatial
  modulation,'' \emph{IEEE IEEE Trans. Veh. Technol.}, vol.~57, no.~4, pp.
  2228--2241, July. 2008.

\bibitem{PLSchallenges2015}
W.~{Trappe}, ``The challenges facing physical layer security,'' \emph{IEEE
  Commun. Mag.}, vol.~53, no.~6, pp. 16--20, June 2015.

\bibitem{Feng2018Two}
F.~{Shu}, Z.~{Wang}, R.~{Chen}, Y.~{Wu}, and J.~{Wang}, ``Two high-performance
  schemes of transmit antenna selection for secure spatial modulation,''
  \emph{IEEE Trans. Veh. Technol.}, vol.~67, no.~9, pp. 8969--8973, Sep. 2018.

\bibitem{shu2018secure}
F.~{Shu}, X.~{Wu}, J.~{Hu}, J.~{Li}, R.~{Chen}, and J.~{Wang}, ``Secure and
  precise wireless transmission for random-subcarrier-selection-based
  directional modulation transmit antenna array,'' \emph{IEEE J. Sel. Areas
  Commun.}, vol.~36, no.~4, pp. 890--904, Apr. 2018.

\bibitem{xia2019antenna}
G.~{Xia}, F.~{Shu}, Y.~{Zhang}, J.~{Wang}, S.~{ten Brink}, and J.~{Speidel},
  ``Antenna selection method of maximizing secrecy rate for green secure
  spatial modulation,'' \emph{IEEE Trans. Green Commun and Netw.}, vol.~3,
  no.~2, pp. 288--301, Jun. 2019.

\bibitem{Wu2015Secret}
F.~{Wu}, L.~{Yang}, W.~{Wang}, and Z.~{Kong}, ``Secret precoding-aided spatial
  modulation,'' \emph{IEEE Commun. Lett.}, vol.~19, no.~9, pp. 1544--1547, Sep.
  2015.

\bibitem{Schmidhuber2015Deep}
J.~Schmidhuber, ``Deep learning in neural networks: An overview,'' \emph{Neural
  Netw}, vol.~61, pp. 85--117, 2015.

\bibitem{C.Jiang2017Machine}
C.~{Jiang}, H.~{Zhang}, Y.~{Ren}, Z.~{Han}, K.~{Chen}, and L.~{Hanzo},
  ``Machine learning paradigms for next-generation wireless networks,''
  \emph{Wireless Commun.}, vol.~24, no.~2, pp. 98--105, Apr. 2017.

\bibitem{He2018Deep}
H.~{He}, C.~{Wen}, S.~{Jin}, and G.~Y. {Li}, ``Deep learning-based channel
  estimation for beamspace mmwave massive mimo systems,'' \emph{IEEE Wireless
  Communications Letters}, vol.~7, no.~5, pp. 852--855, Oct. 2018.

\bibitem{Nachmani2016Learning}
E.~{Nachmani}, Y.~{Be'ery}, and D.~{Burshtein}, ``Learning to decode linear
  codes using deep learning,'' in \emph{2016 54th Annual Allerton Conference on
  Communication, Control, and Computing (Allerton)}, Sep. 2016, pp. 341--346.

\bibitem{Hongji}
H.~Huang, J.~Yang, H.~Huang, Y.~Song, and G.~Gui, ``Deep learning for
  super-resolution channel estimation and {DOA} estimation based massive {MIMO}
  system,'' \emph{IEEE Trans. Veh. Technol.}, vol.~67, no.~9, pp. 8549--8560,
  Aug. 2018.

\bibitem{Kim2018Deep}
M.~{Kim}, N.~{Kim}, W.~{Lee}, and D.~{Cho}, ``Deep learning-aided scma,''
  \emph{IEEE Commun. Lett.}, vol.~22, no.~4, pp. 720--723, Apr. 2018.

\bibitem{nachmani2018deep}
E.~{Nachmani}, E.~{Marciano}, L.~{Lugosch}, W.~J. {Gross}, D.~{Burshtein}, and
  Y.~{Be¡¯ery}, ``Deep learning methods for improved decoding of linear
  codes,'' \emph{IEEE J. Sel. Topics Signal Process.}, vol.~12, no.~1, pp.
  119--131, Feb. 2018.

\bibitem{shu2019high}
F.~{Shu}, X.~{Liu}, G.~{Xia}, T.~{Xu}, J.~{Li}, and J.~{Wang},
  ``High-performance power allocation strategies for secure spatial
  modulation,'' \emph{IEEE Trans. Veh. Technol.}, vol.~68, no.~5, pp.
  5164--5168, May 2019.

\end{thebibliography}
\bibliographystyle{IEEEtran}
\begin{IEEEbiographynophoto}\\
FENG SHU is a professor with the School of Electronic and Optical Engineering at Nanjing University of Science and Technology, Nanjing, China.   He is also with the College of
Physics and Information, Fuzhou University, Fuzhou 350116, China, and the College of Computer and  information at Fujian Agriculture and Forestry University, Fuzhou 350002, China. He has been awarded with Mingjian Scholar Chair Professor and Fujian Hundred-talents Program in Fujian Province, China.   He has published more 200 journal  papers on signal processing and
communications, with more than 100 SCI-indexed papers and more than 80 IEEE journal papers.  Her research interests include  MIMO and beamforming technologies, machine learning for mobile communications£¬ and multiple access techniques.
\end{IEEEbiographynophoto}
\begin{IEEEbiographynophoto}\\
LIN LIU is a postgraduate student in the School of Electronic and Optical Engineering at Nanjing University of Science and Technology, Nanjing, China. His research interests include massive MIMO, wireless localization, and DOA measurement in wireless communications.
\end{IEEEbiographynophoto}
\begin{IEEEbiographynophoto}\\
 YUMENG ZHANG is a graduate student in the School of Electronic and Optical Engineering at Nanjing University of Science and Technology, Nanjing, China. Her research interests include massive MIMO, wireless localization, and DOA measurement in wireless communications.
\end{IEEEbiographynophoto}
\begin{IEEEbiographynophoto}\\
 GuiYANG Xia is a PhD student in the School of Electronic and Optical Engineering at Nanjing University of Science and Technology, Nanjing, China. His research interests include massive MIMO, spatial modulation, and DOA measurement in wireless communications.
\end{IEEEbiographynophoto}
\begin{IEEEbiographynophoto}\\
 XIAOYU LIU is a postgraduate student in the School of Electronic and Optical Engineering at Nanjing University of Science and Technology, Nanjing, China. Her research interests include massive MIMO, wireless localization, and DOA measurement in wireless communications.
\end{IEEEbiographynophoto}
\begin{IEEEbiographynophoto}\\
JUN LI is a professor with the School of Electronic and Optical Engineering at Nanjing University of Science and Technology, Nanjing, China.  His research interests include caching, computing, and channel coding.
\end{IEEEbiographynophoto}
\begin{IEEEbiographynophoto}\\
SHI JIN is a professor School of Information Science and Enginering  at Southeast University, Nanjing, China.  He has published more 100 journal and conference papers on signal processing and communications. His research
interests include space-time wireless communications, random matrix theory, and information theory.
\end{IEEEbiographynophoto}
\begin{IEEEbiographynophoto}\\
JIANGZHOU WANG is currently a Professor and the Former Head of the School of Engineering and Digital Arts, The University of Kent, U.K. He has published over 300 papers in international journals and conferences and four books in the areas of wireless mobile communications.
%is currently the Head of the School of Engineering and Digital Arts and a Professor of Telecommunications with the University of Kent, Canterbury, U.K. He has authored over 200 papers in international journals and conferences in the areas of wireless mobile communications and three books.
His research interests include massive MIMO and beamforming technologies, machine learning for mobile communications and multiple access techniques.
\end{IEEEbiographynophoto}
%\bibliography{IEEEfull,reference}

\end{document}